\documentclass[12pt,authoryear]{article}
\usepackage{amsmath,amsthm}
\usepackage{amssymb,amsfonts,amscd,mathrsfs,latexsym}
\usepackage{mathtools}
\usepackage{thmtools}
\usepackage{graphicx}
\usepackage{color}
\usepackage{xcolor}
\usepackage{natbib}
\usepackage{txfonts}
\usepackage{enumitem}
\usepackage{relsize}
\usepackage{setspace}
\usepackage{accents}
\usepackage{titlesec}
\usepackage{indentfirst}
\usepackage[top=1.75in,bottom=1.25in,left=1.5in,right=1.5in]{geometry}
\usepackage[colorlinks,citecolor=blue,urlcolor=blue,
linkcolor=blue]{hyperref}

\DeclareMathOperator{\gr}{gr}
\DeclareMathOperator{\sgn}{sgn}

\theoremstyle{plain}
\newtheorem{thrm}{Theorem}
\newtheorem*{thrm*}{Theorem}
\newtheorem{lemm}[thrm]{Lemma}
\newtheorem*{lemm*}{Lemma}
\newtheorem{prop}[thrm]{Proposition}
\newtheorem*{prop*}{Proposition}
\newtheorem{corl}[thrm]{Corollary}
\newtheorem*{corl*}{Corollary}

\newtheorem*{claim*}{Claim}

\theoremstyle{definition}
\newtheorem{defn}{Definition}
\newtheorem*{defn*}{Definition}
\newtheorem{rmrk}[thrm]{Remark}
\newtheorem*{rmrk*}{Remark}

\newtheorem{asmp*}{Assumption}

\newtheorem*{exmp*}{Example}

\titleformat{\section}{\fontsize{14}{16}\bfseries}{\thesection}{1em}{}[]
\titleformat{\subsection}{\fontsize{12}{14}\itshape}{\thesubsection}{0.5em}{}[]

\begin{document}

\title{\Large{\textbf{Equilibrium existence in a discrete-time endogenous growth model with physical and human capital}}
}

\author{\large{Luis A. Alcal\'{a}}\thanks{Instituto de Matem\'{a}tica Aplicada San Luis (UNSL and CONICET) and Departamento de Matem\'{a}tica, Universidad Nacional de San Luis, San Luis  Argentina. Email: \href{mailto:lalcala@unsl.edu.ar}{lalcala@unsl.edu.ar}}
}

\date{\large{February 1, 2025}}

\maketitle

\begin{abstract}
This paper studies a discrete-time version of the Lucas-Uzawa endogenous growth model with physical and human capital in the presence of externalities. Existence of an optimal equilibrium is proved using tools from dynamic programming with bounded or unbounded returns. The proofs also rely on properties of isoelastic utility and homogeneous production functions and apply well-known inequalities in real analysis, seldom used in the literature, which significantly simplify the task of verifying certain assumptions that are rather technical in nature. Some advantages of adopting a parametric family of isoelastic utility functions, instead of the ad hoc formulation typically used, are also discussed.
\medskip


\end{abstract}

\section{Introduction}
\label{sec:intro}

Following the tradition of \cite{lucas88} and \cite{uzawa65}, this paper studies a discrete-time model of optimal growth with human capital externalities,  isoelastic utility (that may be unbounded), homogeneous technology that allows for physical and human capital depreciation, and constant population growth. 

Existence of an optimal equilibrium, which includes a characterization of the value function associated with the social planner's problem, is proved using tools from dynamic programming with bounded or unbounded returns, developed by \cite{alvarezstokey98} for the case of homogeneous functions, later unified and generalized in \cite{levanmorhaim02} and \cite{levan06}. This is a direct approach that requires some Lipschitz-type conditions on the primitives of the model.\footnote{There is a vast and ever-expanding literature that followed this work, which was also led by the local contractions method of \cite{boyd90} and \cite{duran00}. Other unifying approaches include \cite{levanvailakis05} for the case of recursive utilities, an analysis of stochastic optimal growth models in \cite{kamihigashi07}, and a generalization of the recursive approach based on temporal aggregators by \cite{bichetal18}, just to give a few examples.}

For clarity of exposition, the main results of the paper are first developed for the model without human capital externalities. It is shown that applying   properties of isoelastic utility and homogeneous production functions, and some inequalities for means, seldom used in the economic literature, significantly simplifies the task of verifying certain technical assumptions. Externalities are included in the model, following two approaches: by a simple  change of variables, which requires minor modifications, and extending the basic framework to show that most results are still valid in a straightforward manner.  

Another relevant aspect, often overlooked, relates to the parametric  family of isoelastic utility functions, which can be derived as the solution to a boundary value problem. These functions have continuity and monotonicity properties that are useful to determine pointwise upper and lower bounds for the optimal solution, also simplifying the proofs and completing the previous analysis of the model with externalities. On the other hand, the alternative ad hoc formulation typically used (obtained by ignoring an additive constant) does not satisfy these properties and requires separate  techniques for three different cases, effectively going against the spirit of a unifying approach.

\subsection{Related models}

Early contributions to the endogenous growth literature were formulated in continuous time. One of the first rigorous treatments of the Lucas-Uzawa model in discrete time is due to \cite{mitra98}. The existence of an equilibrium is proved using a fixed-point theorem by Tychonoff, under certain  simplifying assumptions, such as the absence of physical and human capital depreciation and a bounded utility function, that turn out to be significant. Similar observations apply to the discrete-time version of Romer's model in \cite{levanetal02}.

In papers close to ours, \cite{gourdetal04} and \cite{dalbislevan06} analyze  simple versions of endogenous growth models with externalities, with no  physical capital and bounded returns, to show the existence of optimal and competitive equilibria as solutions of fixed-point problems. Other related work is presented in \cite{hiraguchi11}, where the Lucas-Uzawa model is extended to account for habit formation, but human capital externalities are removed. This paper uses tools from dynamic programming with unbounded returns to show the existence of an interior optimal equilibrium and its convergence to a balanced growth path.

Recent contributions to this literature have focused on establishing conditions to ensure sustained growth. For instance, \cite{hahuytran20} analyze them in a fairly general setting, but their results apply to a one-dimensional state space. More recently, \cite{bosietal23} study a simplified endogenous growth model with logarithmic utility and a linear technology for human capital accumulation. Assuming that the value function is supermodular, they show that optimal paths are either constant or strictly monotonic, which implies the possibility of sustained positive and negative growth (``growth and degrowth'') in their model. Although this is not the focus of the current work, it is shown below that both possibilities may also arise.

\subsection{Organization}

The rest of the paper is organized as follows. Section \ref{sec:model} presents the basic model. The main results of the paper are developed in Section \ref{sec:equilibrium} for the model without human capital externalities. Section \ref{sec:externality} shows that most results are still valid if the basic framework is extended to include  externalities \`{a} la \cite{lucas88}. Section \ref{sec:discuss} discusses the importance of certain assumptions, in particular the use of a family of isoelastic utility functions, in simplifying technical aspects of the proofs. Section \ref{sec:conclusion} offers concluding remarks and directions for future research.

Appendix \ref{sec:inequalities} reviews some of the basics of weighted generalized or power mean inequalities. In particular, the weighted geometric-arithmetic inequality, a result that is used in several proofs. The family of isoelastic utility functions is obtained as the solution of a boundary value problem in Appendix \ref{sec:apputility}, together with the proof of a useful monotonicity property satisfied by these functions.

\section{A two-sector model of endogenous growth}
\label{sec:model}

Time is discrete and denoted by $t \in \mathbb{Z}_+$. The economy is populated by a large number of identical, infinitely-lived agents with unit mass. In each period $t$, there is a single good $y_t$ that is produced using two inputs: physical capital $k_t \in \mathbb{R}_+$, and human capital $h_t \in \mathbb{R}_+$. These inputs depreciate every period at constant rates, given by $0 < \delta_k < 1$ and $0 < \delta_h < 1$, respectively. Each agent has an endowment of one time unit per period, from which $u_t \geq 0$ is allocated to market activities and $v_t \geq 0$ to human capital accumulation. Output per capita is produced with a Cobb-Douglas technology, which is given by $y_t = A k_t^\alpha \left(u_t h_t\right)^{1-\alpha}$ with $A > 0$ and $0 < \alpha < 1$. Let $n \geq 0$ denote the exogenous population growth rate. 

The per capita aggregate resource restriction is then
\begin{align*}
c_t + (1+n)k_{t+1} \leq A k_t^\alpha \left(u_t h_t\right)^{1-\alpha}+ (1-\delta_k)k_t, & & t=0,1,\ldots
\end{align*}
Human capital accumulation is linear in $h_t$ and given by
\begin{align*}
h_{t+1} =\left[B\phi(v_t) + (1-\delta_h)\right]h_t, & & t=0,1,\ldots, 
\end{align*}
where $B > 0$ and $\phi:[0,1] \to \mathbb{R}_+$ satisfies the following assumptions:
\smallskip
\begin{enumerate}[label=(H\arabic*),itemsep=0pt]
\item\label{itm:H1} $\phi$ is continuous and strictly increasing on $[0,1]$;
\item\label{itm:H2} $\phi(0)=0$, and $\phi(1) > \delta_h/B$.
\end{enumerate} 
Note that \cite{lucas88} uses $\phi(v)=v$, while \cite{uzawa65} assumes that $\phi$ is increasing and concave, but both authors ignore human capital depreciation. The strict monotonicity in \ref{itm:H1} is assumed for simplicity and it may be weakened at the expense of clarity, without affecting significantly the results. \ref{itm:H2} ensures that growth rates are bounded above and the existence of constant paths in the model. For ease of notation, the maximum growth rate for human capital in a given period is denoted as
$D_h:=B\phi(1) + (1-\delta_h)$. 

The instantaneous utility of a representative agent $U:\mathbb{R}_+ \!\to \mathbb{R} \cup \{-\infty\}$ belongs to the family of isoelastic utility functions:
\begin{equation}
\label{eq:Udef}
U(c)=
\begin{cases}
\dfrac{c^\theta-1}{\theta}, & \text{if }\ -\infty < \theta \leq 1,\ \theta \neq 0,\\[7pt]
\log{c}, & \text{if }\ \theta = 0.
\end{cases}
\end{equation}
Let $\beta \in (0,1)$ be the discount factor. Along an optimal path, a social planner chooses a sequence $\left\{\left(c_t,u_t,v_t,k_{t+1},h_{t+1}\right)\right\}_{t=0}^{+\infty}$ that solves the following problem
\begin{align}
\label{eq:SP} 
\sup\ \ &\sum_{t=0}^{+\infty} \beta^t U(c_t)\\[3pt]
\label{eq:RCSP}
\text{s.t.}\ \ 
&c_t + (1+n)k_{t+1} \leq A k_t^\alpha \left(u_t h_t\right)^{1-\alpha}+ (1-\delta_k)k_t,& & t=0,1,\ldots,\\[3pt]
\label{eq:HCSP}
& h_{t+1} =\left[B\phi(v_t) + (1-\delta_h)\right]h_t,& & t=0,1,\ldots,\\[3pt] 
\label{eq:TCSP}
& u_t + v_t \leq 1,& & t=0,1,\ldots,\\[3pt]
\label{eq:NCSP}
& c_t,u_t,v_t,k_{t+1},h_{t+1} \geq 0,& & t=0,1,\ldots,
\\[3pt]
& (k_0,h_0) \in \mathbb{R}_+^2\ \text{given}. \nonumber
\end{align}
A solution to this dynamic program, if it exists, is called an \textbf{optimal equilibrium} for the model without externalities. 

Since the objective of this optimization problem is strictly monotone in each variable $c_t$, $u_t$, and $v_t$, the resource constraint \eqref{eq:RCSP} and time constraint \eqref{eq:TCSP} in the planner's problem will hold with equality. Taking these facts into consideration, problem \eqref{eq:SP} can be reformulated in terms of the choice of $(k_{t+1},h_{t+1})$ only, for all $t \geq 0$. As the following lemma shows, there is a \emph{minimum} fraction of time $\underline{v}$ in the human capital sector needed for this economy to exhibit sustained growth along a stationary path. This result will be useful to reformulate the problem and for later proofs. 

\begin{lemm}
\label{lem:vbar}
There exists a unique value $\underline{v} \in (0,1)$ such that
$B\phi(v) + (1-\delta_h) \leq 1$, for all $v \in [0,\underline{v}]$, and $B\phi(v) + (1-\delta_h) > 1$, for all $v \in (\underline{v},1]$.
\end{lemm}

\begin{proof}
By assumption \ref{itm:H2}, we have that 
\begin{align*}
B\phi(0)+(1-\delta_h)=(1-\delta_h) < 1 \quad \text{and} 
\quad B\phi(1)+(1-\delta_h) > 1.
\end{align*}
Then, the intermediate value theorem implies there exists $\underline{v} \in (0,1)$, implicitly defined by $\phi(\underline{v})=\delta_h/B$, such that $B\phi(\underline{v})+(1-\delta_h)=1$. From \ref{itm:H1}, $\phi$ is strictly increasing, hence $\underline{v}$ must be unique. The remaining properties follow from the continuity of $\phi$.     
\end{proof}

Given that \eqref{eq:TCSP} holds with equality, substituting $v_t = 1 - u_t$ into \eqref{eq:HCSP} allows to define a function $\psi:\mathbb{R}_{+}^2 \to [0,1]$ by
\begin{align}
\label{eq:psidef}
\psi\left(h_t,h_{t+1}\right)
=
\begin{cases}
1, &0 \leq h_{t+1} \leq (1-\delta_h)h_t,\\[5pt]
1-\phi^{-1}\left[\frac{1}{B}\left(\frac{h_{t+1}}{h_t}-(1-\delta_h)\right)\right], &(1-\delta_h)h_t \leq h_{t+1} \leq D_h h_t,
\end{cases}
\end{align}
for all $h_t > 0$, and $\psi(0,h_{t+1})=1$ for all $h_{t+1} \geq 0$. It follows directly from the definition in \eqref{eq:psidef} and Lemma \ref{lem:vbar} that there is a \emph{maximum} fraction of time devoted to market activities $\overline{u}$ compatible with sustained growth if the system  follows a stationary path, which is defined by $\overline{u}:=\psi(h,h)=1-\phi^{-1}(\delta_h/B)$, for all $h > 0$. In each period $t$ where $u_t = \overline{u}$ and $h_t>0$, human capital is kept constant at its current level, i.e., $h_{t+1} = h_t$, and, for any $0 \leq u_t < \overline{u}$, human capital increases next period, i.e., $h_{t+1} > h_t$. 

Let $(k_t,h_t) \in \mathbb{R}_+^2$. Combining restrictions \eqref{eq:TCSP} and \eqref{eq:NCSP}, for $c_t$, $u_t$, and $v_t$, with \eqref{eq:RCSP} and \eqref{eq:HCSP} yields the feasible choices for the state next period $(k_{t+1},h_{t+1})$ as
\begin{align}
\label{eq:RRk}
0 &\leq k_{t+1} \leq \frac{Ak_t^\alpha h_t^{1-\alpha}+(1-\delta_k)k_t}{(1+n)},\\[5pt]
\label{eq:RRh}
0 &\leq h_{t+1} \leq D_h h_t,
\end{align}
for each $t=0,1,\ldots$ These two conditions define a feasibility correspondence $\Gamma: \mathbb{R}_+^2 \to \mathbb{R}_+^2$, given by
\begin{align}
\label{eq:Gammadef}
\Gamma(k_t,h_t):=\left\{(k_{t+1},h_{t+1})\in\mathbb{R}_+^2:\eqref{eq:RRk}\ \text{and}\ \eqref{eq:RRh}\ \text{hold for some}\ (k_t,h_t) \in \mathbb{R}_+^2\right\}.
\end{align}
Hence, the social planner's problem \eqref{eq:SP} can be reformulated as
\begin{align}
\label{eq:SPref}
\sup_{\{(k_{t+1},h_{t+1})\}_{t=0}^{+\infty}}\ &\sum_{t=0}^{+\infty} \beta^t F\left(k_t,h_t,k_{t+1},h_{t+1}\right)\\[5pt]
\text{s.t.}\quad &(k_{t+1},h_{t+1}) \in \Gamma(k_t,h_t), & t=0,1,\ldots,\nonumber\\[5pt]
& (k_0, h_0) \in \mathbb{R}_+^2\ \text{given},\nonumber
\end{align}
where the return function $F:\mathbb{R}_+^2 \times \mathbb{R}_+^2 \to \mathbb{R} \cup \{-\infty\}$ is defined by
\begin{align}
\label{eq:Fdef}
F(k_t,h_t,k_{t+1},h_{t+1}) := U\left[Ak_t^\alpha
\left(\psi(h_t,h_{t+1})h_t\right)^{1-\alpha}+(1-\delta_k)k_t-(1+n)k_{t+1}\right],
\end{align}
for all $(k_t,h_t,k_{t+1},h_{t+1}) \in \mathbb{R}_{++}^2\!\times \mathbb{R}_+^2$ for which $U(\cdot)> -\infty$. In the cases where any of $k_t$ or $h_t$ is zero, 
\begin{align*}
 F(k_t,h_t,k_{t+1},h_{t+1}):=U\left[\max\left\{0,(1-\delta_k)k_t-(1+n)k_{t+1}\right\}\right],
\end{align*}
for all $(k_{t+1},h_{t+1}) \in \mathbb{R}_+^2$.

Finally, for any $(k_0,h_0) \in \mathbb{R}_+^2$, let
\begin{align}
\label{eq:Pidef}
\Pi(k_0,h_0):= \left\{\{(k_t,h_t)\}_{t=0}^{+\infty}:(k_{t+1},h_{t+1}) \in \Gamma(k_t,h_t),\ t=0,1,\ldots\right\}
\end{align}
be the set of all sequences of the state variables that are feasible from $(k_0,h_0)$.  A typical element of $\Pi$ will be denoted by $(\mathbf{k},\mathbf{h})$.

\section{Equilibrium existence and characterization}
\label{sec:equilibrium}

The analysis of equilibrium existence for the model developed in the previous section is mainly  based on \cite{levanmorhaim02} and \cite{levan06}. For ease of notation, the norm of any vector $(k,h) \in \mathbb{R}_+^2$ is written as $\|k,h\|$. If the current state of the system is given by $(k,h)$, the next period value is represented by $(k',h')$. Also, the graph of $\Gamma$ is denoted as $\gr(\Gamma)$.

A number of assumptions on $\Gamma$, $F$, and $\beta$  will be used to state and prove the results.
\begin{enumerate}[label=(A\arabic*),itemsep=0pt,topsep=2pt]
\item\label{itm:A1} $\Gamma$ is a nonempty, continuous, compact-valued correspondence. Moreover, $(0,0) \in \Gamma(0,0)$.
\item\label{itm:A2} There exist constants $\zeta \geq 0$ and $\zeta' \geq 0$, with $\zeta \neq 1$, such that $(k',h') \in \Gamma(k,h)$ implies  
$\|k',h'\| \leq \zeta\|k,h\| + \zeta'$. 
\item\label{itm:A3} There exist constants $\eta \geq 0$ and $\eta' \geq 0$, such that 
\begin{align*}
F(k,h,k',h') \leq \eta \left(\|k,h\|+\|k',h'\|\right) + \eta',
\end{align*} 
for all $(k,h,k',h') \in \gr(\Gamma)$.
\item\label{itm:A4} If $\zeta > 0$ in \ref{itm:A2}, then $\beta\zeta < 1$.
\item\label{itm:A5} $F$ is continuous at any point in $\gr(\Gamma)$ such that $F(k,h,k',h') > -\infty$. If $F(k,h,k',h') = -\infty$, then for any sequence $\left\{(k_n,h_n,k_n',h_n')\right\}_{n=0}^{+\infty}$ in $\gr(\Gamma)$ that converges to $(k,h,k',h')$, it follows that 
$\displaystyle\lim_{n \to +\infty} F(k_n,h_n,k_n',h_n')= -\infty$.  
\item\label{itm:A6} For all $(k,h,k',h') \in \gr(\Gamma)$ and for all $\lambda \in (0,1]$,
\begin{enumerate}
\item $(\lambda k,\lambda h,\lambda k',\lambda h') \in \gr(\Gamma)$;
\item there exist continuous functions $\Phi_1:(0,1]\to\mathbb{R}$ and $\Phi_2:(0,1]\to\mathbb{R}$, with $\Phi_1(1)=1$ and $\Phi_2(1)=0$, that satisfy 
\begin{align*}
F(\lambda k,\lambda h,\lambda k',\lambda h') \geq \Phi_1(\lambda) F(k,h,k',h') + \Phi_2(\lambda);
\end{align*}
\item if $(\tilde{k},\tilde{h}) \in \mathbb{R}_+^2$, then for all $(k',h') \in \Gamma(\tilde{k},\tilde{h})$ and $\varepsilon > 0$ sufficiently small, there exists a neighborhood $\mathcal{N}$ of $(\tilde{k},\tilde{h})$ in $\mathbb{R}_+^2$, such that 
\begin{align*}
\left((1-\varepsilon)k',(1-\varepsilon)h'\right) \in \Gamma(k,h),\quad \text{for each}\ (k,h) \in \mathcal{N}(\tilde{k},\tilde{h}).   
\end{align*}
\end{enumerate} 
\end{enumerate}

\begin{lemm}
\label{lem:Pidef}
Assume that \ref{itm:A1} and \ref{itm:A2} are satisfied. Then 
\begin{enumerate}[label=\emph{(\alph*)},itemsep=0pt]
\item $\Pi(k_0,h_0)$ is compact in the product topology, for all $(k_0,h_0) \in \mathbb{R}_+^2$;
\item The mapping $\Pi:\mathbb{R}_+^2 \to \big(\mathbb{R}_+^2\big)^\infty$ is continuous in the product topology.
\end{enumerate}
\end{lemm}

\begin{proof}
First, it is shown that \ref{itm:A1} and \ref{itm:A2} are verified. From the definition given in \eqref{eq:Gammadef}, it is clear that $(0,0) \in \Gamma(0,0)$. In fact, $\Gamma$ satisfies the stronger condition $\Gamma(0,0)=\{(0,0)\}$ in this case. For each $(k,h) \in \mathbb{R}_+^2$, the feasibility correspondence can be written, by \eqref{eq:RRk} and \eqref{eq:RRh}, as the Cartesian product of two closed intervals:  
\begin{align*}
\left[0,(1+n)^{-1}(Ak^\alpha h^{1-\alpha}+(1-\delta_k)k)\right] \quad \text{and} \quad \left[0,D_h h\right],
\end{align*}
which is a compact set. Thus, $\Gamma$ is compact-valued. The continuity of $\Gamma$ can be proved with standard arguments, hence it is omitted. Therefore, \ref{itm:A1} is satisfied.

Now suppose that $(k',h') \in \Gamma(k,h)$ for some $k,h \geq 0$. Applying the weighted geometric-arithmetic mean inequality,\footnote{See   Appendix \ref{sec:inequalities} for details.} or weighted GM-AM inequality, to the right-hand side of \eqref{eq:RRk} yields
\begin{align*}
k' \leq \frac{A k^\alpha h^{1-\alpha}+(1-\delta_k)k}{(1+n)}   
&\leq \frac{A\left[\alpha k + (1-\alpha)h\right] + (1-\delta_k)k}{(1+n)},\\[5pt]
&= \frac{\left[\alpha A + (1-\delta_k)\right]k + (1-\alpha)A h}{(1+n)}.
\end{align*}
It immediately follows that $k' \leq \xi\|k,h\|$, where 
\begin{equation}
\label{eq:xidef}
\xi := \max\left\{\frac{\alpha A +(1-\delta_k)}{1+n},\frac{(1-\alpha)A}{1+n}\right\}.
\end{equation}
On the other hand, by \eqref{eq:HCSP}, we have that 
\begin{align*}
h' \leq \left[B\phi(1)+(1-\delta_h)\right]\|k,h\| = D_h \|k,h\|.
\end{align*}
Set $\zeta$ to be the greater between $\xi$ and $D_h$, and let $\zeta' = 0$. Clearly, $\zeta > 0$. The condition $\zeta \neq 1$ is also satisfied, for if $\zeta =1$ would contradict that $D_h = B\phi(1)+(1-\delta_h)>1$. Hence, \ref{itm:A2} holds.

The result in part (a) of this Lemma follows from Tychonoff's theorem. The proof for (b) can be found in Lemma 2 on \cite{levanmorhaim02}.
\end{proof}

\begin{lemm}
\label{lem:Jdef}
Suppose that \ref{itm:A1}, \ref{itm:A2}, \ref{itm:A3}, and \ref{itm:A4} are satisfied. Then, the infinite sum in the social planner's problem \eqref{eq:SPref} is well defined and equals 
\begin{align*}
\lim_{T \to +\infty}\ \sum_{t=0}^T \beta^t F(k_t,h_t,k_{t+1},h_{t+1}),
\end{align*}
for all $(\mathbf{k},\mathbf{h}) \in \Pi(k_0,h_0)$ and for all $(k_0,h_0) \in \mathbb{R}_+^2$.
\end{lemm}

\begin{proof}
Since the validity of \ref{itm:A1} and \ref{itm:A2} has already been established in the proof of Lemma \ref{lem:Pidef}, it will suffice to show that \ref{itm:A3} and \ref{itm:A4} also hold. A detailed proof of the result for a general case is given in \cite{levan06} (Lemma 2.1.1 and Remark 2.2.1). 

It is easily verified from \eqref{eq:Udef} that $U$ is pointwise bounded on $\mathbb{R}_+$. In particular, as $\theta \to 1^-$, it follows that $U(c) \leq c-1$, for all $c \geq 0$. From this fact, and given that $\psi(h,h') \leq 1$ for all $(h,h') \in \mathbb{R}_+^2$, we have from \eqref{eq:Fdef} that 
\begin{align*}
F(k,h,k',h') &\leq Ak^\alpha\left(\psi(h,h')h\right)^{1-\alpha}+(1-\delta_k)k-(1+n)k'-1,\\[3pt]
&\leq Ak^\alpha h^{1-\alpha}+(1-\delta_k)k-(1+n)k'-1,
\end{align*}
for every $(k,h,k',h') \in \gr(\Gamma)$. Therefore,
\begin{align*}
F(k,h,k',h') &\leq A\left[\alpha k +(1-\alpha) h\right] + (1-\delta_k)k - (1+n)k' - 1,\\[5pt]
&\leq \left[\alpha A+(1-\delta_k)\right] k + (1-\alpha)A h - (1+n)k' - 1,\\[5pt]
&\leq \zeta(1+n)\|k,h\|,\\[5pt]
&\leq \zeta(1+n)\left(\|k,h\| + \|k',h'\|\right),
\end{align*}
where the weighted GM-AM inequality \eqref{eq:WAMGMI} is applied on the first line, as in the proof of the previous lemma. Thus, \ref{itm:A3} is satisfied with $\eta:=\zeta(1+n)$ and $\eta':=0$.

In order for \ref{itm:A4} to hold, we simply impose the condition $\beta\zeta < 1$, which is equivalent to
\begin{align}
\label{eq:betacond}
\beta\cdot\max\left\{\frac{\alpha A +(1-\delta_k)}{(1+n)},\frac{(1-\alpha)A}{(1+n)},D_h\right\} < 1.
\end{align}
This completes the proof.
\end{proof}

Lemma \ref{lem:Jdef} allows to define total discounted returns  $J:\Pi \to \mathbb{R} \cup \{-\infty\}$ as 
\begin{align}
\label{eq:Jdef}
J(\mathbf{k},\mathbf{h}):=\sum_{t=0}^{+\infty} \beta^t F(k_t,h_t,k_{t+1},h_{t+1}). 
\end{align} 
Then, problem \eqref{eq:SPref} can be written more compactly as
\begin{align}
\label{eq:PPcompact}
\sup\ \left\{J(\mathbf{k},\mathbf{h}):(\mathbf{k},\mathbf{h}) \in \Pi(k_0,h_0)\right\}.
\end{align} 

\begin{prop}
\label{prp:exist}
Assume that \ref{itm:A1}, \ref{itm:A2}, \ref{itm:A3}, and \ref{itm:A4} are satisfied. Then, a solution to problem \eqref{eq:PPcompact} exists.
\end{prop}

\begin{proof}
Applying the result from Lemma 2.2.1 in \cite{levan06}, the function $J(\mathbf{k},\mathbf{h})$ given in \eqref{eq:Jdef} is upper semi-continuous in the product topology for each $(\mathbf{k},\mathbf{h}) \in \Pi(k_0,h_0)$. Since $\Pi(k_0,h_0)$ is compact in the same topology, by Lemma \ref{lem:Pidef}, then $J$ attains its maximum on  $\Pi(k_0,h_0)$.
\end{proof}

In order to further characterize the solution, it is useful to define the set of feasible paths for which total discounted returns are finite, that is, 
\begin{align*}
\Pi'(k_0,h_0):=\left\{(\mathbf{k},\mathbf{h}) \in \Pi(k_0,h_0): J(\mathbf{k},\mathbf{h}) > -\infty\right\}.
\end{align*}
The following lemma, related to this set, is important for later proofs.

\begin{lemm}
\label{lem:pitilde}
$\Pi'(k_0,h_0)$ is nonempty for all $(k_0,h_0) \in \mathbb{R}_{++}^2$.
\end{lemm}

\begin{proof}
If $0 < \theta \leq 1$ in \eqref{eq:Udef}, then $F$ is bounded below by $F(0,0,0,0)=-\theta^{-1} > -\infty$, so the result is immediate. Some analysis is required for the case $-\infty < \theta \leq 0$. Let $(k_0,h_0) \gg 0$. Assume that  $u_t = \overline{u}$, for all $t \geq 0$, and 
\begin{align}
\label{eq:k0low}
k_0 < \left(\frac{A}{n+\delta_k}\right)^{\frac{1}{1-\alpha}}\overline{u}h_0.
\end{align}
By Lemma \ref{lem:vbar} and the definition of $\overline{u}$, it follows that $h_t=h_0$, for all $t$. Then, it is possible to choose a constant consumption sequence $\mathbf{c}_0=(c_0,c_0,\ldots)$, with $c_0:= Ak_0^\alpha\left(\overline{u}h_0\right)^{1-\alpha}-(n+\delta_k)k_0 > 0$, such that
\begin{align*}
k_{t+1} = \frac{Ak_0^\alpha\left(\overline{u}h_0\right)^{1-\alpha}+(1-\delta_k)k_0-c_0}{1+n}=k_0, & & \text{for all}\ t=0,1,\ldots
\end{align*}
Clearly, the sequences $\mathbf{k}_0=(k_0,k_0,\ldots)$ and $\mathbf{h}_0=(h_0,h_0,\ldots)$ are in $\Pi(k_0,h_0)$, and
\begin{align*}
J(\mathbf{k}_0,\mathbf{h}_0)
=\frac{U[Ak_0^\alpha\left(\overline{u}h_0\right)^{1-\alpha}-(n+\delta_k)k_0]}{1-\beta} > -\infty.
\end{align*}
If \eqref{eq:k0low} is not satisfied, i.e., $k_0 \geq \left(A/(n+\delta_k)\right)^{\frac{1}{1-\alpha}}\left(\overline{u}h_0\right)$, then set $k_1 >0$ so that 
$c_0 = Ak_0^\alpha\left(\overline{u}h_0\right)^{1-\alpha}+(1-\delta_k)k_0 - (1+n)k_1 > 0$, and  
$c_1 = Ak_1^\alpha\left(\overline{u}h_0\right)^{1-\alpha}-(n+\delta_k)k_1 > 0$,
which in turn requires that $k_1 < \overline{k}$, where 
\begin{align*}
\overline{k}:= \min\left\{\frac{Ak_0^\alpha\left(\overline{u}h_0\right)^{1-\alpha} + (1-\delta_k)k_0}{(1+n)},\left(\frac{A}{n+\delta_k}\right)^{\frac{1}{1-\alpha}}\overline{u}h_0\right\}.
\end{align*}
Note that, by construction, $c_t = c_1$, for all $t \geq 1$, implies $k_t = k_1$, for all $t \geq 1$. Then, there exist sequences $\mathbf{h}_0=(h_0,h_0,h_0,\ldots)$ and $\mathbf{k}_1=(k_0,k_1,k_1,\ldots)$ that are feasible from $(k_0,h_0)$, whose total discounted value is given by
\begin{align*}
J(\mathbf{k}_1,\mathbf{h}_0) &= U\left[Ak_0^\alpha\left(\overline{u}h_0\right)^{1-\alpha}+(1-\delta_k)k_0 - (1+n)k_1\right]\\
\quad &+\sum_{t=1}^{+\infty}\beta^t U\left[Ak_1^\alpha\left(\overline{u}h_0\right)^{1-\alpha}-(n+\delta_k)k_1\right],\\
&=F(k_0,h_0,k_1,h_0)+\frac{\beta}{1-\beta}\,F(k_1,h_0,k_1,h_0) > -\infty.
\end{align*}
This completes the proof.
\end{proof}

Denote by $V(k_0,h_0)$ the value of problem \eqref{eq:PPcompact}. Let $S$ be the space of functions $W:\mathbb{R}_+^2 \to \mathbb{R} \cup \{-\infty\}$ that are upper semicontinuous and satisfy:
\begin{enumerate}[label=(S\arabic*)]
\item for all $(k_0,h_0) \in \mathbb{R}_+^2$ and $(\mathbf{k},\mathbf{h}) \in \Pi(k_0,h_0)$, $\displaystyle\limsup_{t \to +\infty}\ \beta^t W(k_t,h_t) \leq 0$;
\item for all $(k_0,h_0) \in \mathbb{R}_+^2$ such that $\Pi'(k_0,h_0)$ is not empty and $(\mathbf{k},\mathbf{h}) \in \Pi'(k_0,h_0)$, $\displaystyle\lim_{t \to +\infty} \beta^t W(k_t,h_t) = 0$.
\end{enumerate}

\begin{prop}
\label{prp:bellman}
Suppose that \ref{itm:A1}, \ref{itm:A2}, \ref{itm:A3}, \ref{itm:A4}, and \ref{itm:A5} hold.
\begin{enumerate}[label=\emph{(\alph*)},itemsep=0pt]
\item The value function $V:\mathbb{R}_+^2 \to \mathbb{R} \cup \{-\infty\}$ satisfies the following Bellman equation 
 \begin{align}
\label{eq:BE}
V(k,h) &= \sup\left\{F(k,h,k',h')+\beta V(k',h'):(k',h') \in \Gamma(k,h)\right\},
\end{align}
for all $(k,h) \in \mathbb{R}_+^2$.
\item $V$ is the unique solution to the Bellman equation \eqref{eq:BE} in the space $S$.
\item If, in addition, \ref{itm:A6} holds, $V$ is continuous for any $(\mathbf{k},\mathbf{h}) \in \Pi'(k_0,h_0)$, and if $V(k_0,h_0) = - \infty$, then for any sequence $\{(k_n,h_n)\}_{n=0}^{+\infty}$ that converges to $(k_0,h_0)$, if follows that $\displaystyle\lim_{n \to +\infty} V(k_n,h_n)=-\infty$.  
\end{enumerate}
\end{prop}

\begin{proof}
In order to prove parts (a) and (b), it remains to verify that \ref{itm:A5} holds, and the results follow from Theorem 2 in \cite{levanmorhaim02}. Standard continuity arguments apply, hence the proof is omitted. Next we show that \ref{itm:A6} is satisfied. 

Note from \eqref{eq:psidef} that $\psi$ is homogeneous of degree zero in $(h,h')$. Moreover, \eqref{eq:RRk} and \eqref{eq:RRh} imply that if $(k',h') \in \Gamma(k,h)$, then $(\lambda k',\lambda h') \in \Gamma(\lambda k,\lambda h)$, for every $\lambda \geq 0$; that is, $\Gamma$ is a cone. Hence, \ref{itm:A6}(a) holds. Let $f:\gr(\Gamma)\to \mathbb{R} \cup \{-\infty\}$ be given by \begin{align*}
f(k,h,k',h'):= Ak^\alpha(\psi(h,h')h)^{1-\alpha}+(1-\delta_k)k-(1+n)k',
\end{align*}
so that $F = U \circ f$. Clearly, $f$ is homogeneous of degree one in $(k,h,k',h')$, i.e., $f(\lambda k,\lambda h,\lambda k',\lambda h')=\lambda f(k,h,k',h')$, for all $\lambda \geq 0$, and for all $(k,h,k',h') \in \gr(\Gamma)$. By the definition of $U$ in \eqref{eq:Udef}, it follows that for every $c > 0$ and $\lambda > 0$,
\begin{align}
\label{eq:Ulambdac}
U(\lambda c) = \frac{(\lambda c)^\theta-1}{\theta}=\frac{\lambda^\theta(c^\theta-1)}{\theta}+\frac{\lambda^\theta-1}{\theta}
=\lambda^\theta U(c)+U(\lambda).
\end{align}   
Let $\Phi_1(\lambda):=\lambda^\theta$ and $\Phi_2(\lambda):=U(\lambda)$, for all $\lambda \in (0,1]$, which are both continuous functions. From the fact that $f$ is linearly homogeneous and \eqref{eq:Ulambdac}, 
\begin{align*}
F(\lambda k,\lambda h,\lambda k',\lambda h') = \Phi_1(\lambda)F(k,h,k',h') + \Phi_2(\lambda),
\end{align*}
with $\Phi_1(1)=1$ and $\Phi_2(1)=0$, for all $(k,h,k',h') \in \gr{\Gamma}$, which also implies that \ref{itm:A6}(b) holds. Now, assume that $(k,h) \gg (0,0)$. Therefore
\begin{align*}
0 < k' \leq \frac{Ak^\alpha h^{1-\alpha} +(1-\delta_k)k}{(1+n)} \quad \text{and} \quad 
0 < h' \leq D_h h.
\end{align*}
Moreover, there exists $\varepsilon > 0$ such that
\begin{align*}
0 < (1-\varepsilon)k' < \frac{Ak^\alpha h^{1-\alpha} +(1-\delta_k)k}{(1+n)} \quad \text{and} \quad 0 < (1-\varepsilon)h' < D_h h.
\end{align*}
Since these lower and upper bounds are continuous functions of $k$ and $h$, the above inequalities also hold for every $(\tilde{k},\tilde{h}) \in \mathcal{N}(k,h)$. If $k \geq 0$, $h \geq 0$, then 
\begin{align*}
0 = k' < \frac{Ak^\alpha h^{1-\alpha} +(1-\delta_k)k}{(1+n)} \quad \text{and} \quad 0 = h' < D_h h.
\end{align*}
Simply multiply both sides of the left-hand side equalities by $(1-\varepsilon)$ and both conditions are also verified for all $(\tilde{k},\tilde{h}) \in \mathcal{N}(k,h)$, Then, \ref{itm:A6}(c) is satisfied, and part (c) of the proposition follows from Proposition 1 and Theorem 3(ii) in \cite{levanmorhaim02}.   
\end{proof}

\begin{rmrk}
Note that every result in Proposition \ref{prp:bellman} remains valid if $\theta = 0$, i.e., $U(c)=\log{c}$. In particular, it follows that for all $\lambda \in (0,1]$,
\begin{align*}
\log(\lambda c) = \log{c} + \log{\lambda} = \lambda^0 \log{c}+\log{\lambda},
\end{align*}
thus \eqref{eq:Ulambdac} is satisfied. This in turn implies that $\Phi_1(\lambda)=1$ and $\Phi_2(\lambda)=\log{\lambda}$. Hence, a separate treatment for the case of logarithmic utility, as in \cite{alvarezstokey98} or some of the applications analyzed in \cite{levanmorhaim02} and \cite{levan06},  is not needed with our assumptions.
\end{rmrk}

\section{The model with externalities}
\label{sec:externality}

This section extends the basic model from Section \ref{sec:model} to include human capital externalities \`a la \cite{lucas88}. Let $h_{at}$ denote the average level of human capital in the population. Now output technology is given by $y_{at} = A k_t^\alpha \left(u_t h_t\right)^{1-\alpha}(h_{at})^\gamma$, with $\gamma \geq 0$. In this case, the social planner's problem is similar to the model without externalities, but it requires an additional equilibrium condition. Given that agents are identical with unit mass, it must be the case that $h_{at} = h_t$, for all $t \geq 0$, along any optimal path. A solution to this extended version of the problem is called an \textbf{optimal equilibrium with externalities}.\footnote{In this approach, the social planner ``internalizes the externality'' to choose an optimal plan. Alternatively, this equilibrium can be defined as a fixed point of an appropriate operator, as in \cite{gourdetal04} and \cite{bosietal23}.}

In the presence of human capital externalities, the resource constraint \eqref{eq:RCSP} now takes the form
\begin{align}
\label{eq:RCex}
c_t +(1+n)k_{t+1} \leq A k_t^\alpha u_t^{1-\alpha} h_t^{1-\alpha+\gamma} + (1-\delta_k)k_t, & & t=0,1,\ldots,
\end{align}
and the feasibility restriction \eqref{eq:RRk} is replaced by
\begin{align}
\label{eq:RRkex}
0 \leq k_{t+1} \leq \frac{Ak_t^\alpha h_t^{1-\alpha+\gamma}+(1-\delta_k)k_t}{(1+n)}, & & t=0,1,\ldots,
\end{align}
while the restriction \eqref{eq:RRh} for $h_{t+1}$ remains unchanged.

Two different approaches are considered: (i) making a change of variables, which allows to apply the same framework developed for the basic model, with minor modifications; and (ii) including the externalities explicitly in the model and showing that most results can be extended in a straightforward manner.

\subsection{A change of variables}
\label{sec:changevar}

First, note that the right-hand side of \eqref{eq:RCex} is no longer homogeneous of degree one in $(k_t,h_t)$ if $\gamma >0$.\footnote{It is convenient to allow $\gamma = 0$, which reduces the model to the case without externalities, for ease of comparison.} However, for the sake of proving equilibrium existence, it is possible to maintain linear homogeneity and apply the results from the previous section, by making a simple change of variables. Let 
\begin{align}
\label{eq:hhatdef}
\hat{h}_t:= h_t^{\frac{1-\alpha+\gamma}{1-\alpha}}.
\end{align} 
Now output technology is homogeneous of degree one in $k_t$ and the transformed variable $\hat{h}_t$, i.e., $\hat{y}_t = A k_t^\alpha (u_t\hat{h}_t)^{1-\alpha}$. Furthermore, the restriction \eqref{eq:RRkex} expressed in terms of $\hat{h}_t$ coincides with the corresponding restriction for the model without externalities, given in \eqref{eq:RRk}. Next, substitute \eqref{eq:hhatdef} into \eqref{eq:HCSP} to obtain
\begin{align}
\label{eq:hhatprime}
\hat{h}_{t+1} = \left[B\phi(v_t)+(1-\delta_h)\right]^\rho\hat{h}_t,
\end{align}
where $1 \leq \rho < +\infty$ is given by 
$\rho:=(1-\alpha+\gamma)/(1-\alpha)$.
Hence, human capital accumulation is still linear in $\hat{h}_t$, and strictly increasing in $v_t$ for all $\hat{h}_t > 0$. 
Finally, define a function $\psi_\rho:\mathbb{R}_+^2 \to [0,1]$, to play the same role as $\psi$ in the problem without externalities, by
\begin{align}
\label{eq:psirhodef}
\psi_\rho(\hat{h}_t,\hat{h}_{t+1}) =
\begin{cases}
1, &0 \leq \hat{h}_{t+1} \leq d_h\hat{h}_t,\\[5pt]
1-\phi^{-1}\left\{\frac{1}{B}\left[\left(\frac{\hat{h}_{t+1}}{\hat{h}_t}\right)^\rho-(1-\delta_h)\right]\right\}, &d_h\hat{h}_t \leq \hat{h}_{t+1} \leq D_h^{\,\rho}\hat{h}_t,
\end{cases}
\end{align}
where $D_h^{\,\rho}=\left[B\phi(1)+ (1-\delta_h)\right]^\rho$ represents the maximum attainable growth rate of human capital in one period. Clearly, $D_h^{\,\rho} \geq D_h$, from \ref{itm:H2}.  

Using \eqref{eq:hhatprime}, \eqref{eq:psirhodef}, and the appropriate definitions to obtain expressions for the feasibility correspondence $\Gamma_\rho(k_t,\hat{h}_t)$ and the return function  $F_\rho(k_t,\hat{h}_t,k_{t+1},\hat{h}_{t+1})$, it is easy to verify that all the results given in Section \ref{sec:equilibrium} apply to the social planner's problem \eqref{eq:SPref} formulated for $\Gamma_\rho$, $F_\rho$, and $\beta$, in terms  of the transformed variables.

\subsection{An extension}
\label{sec:extension}

A different, and perhaps more elegant, approach consists in showing that the basic framework applies to the extended model, taking constraints \eqref{eq:RCex} and \eqref{eq:RRkex} into account. Now, the feasibility correspondence $\hat{\Gamma}:\mathbb{R}_+^2 \to \mathbb{R}_+^2$ is given by 
\begin{align}
\label{eq:Gammaexdef}
\hat{\Gamma}(k_t,h_t):=\left\{(k_{t+1},h_{t+1})\in\mathbb{R}_+^2:\eqref{eq:RRkex}\ \text{and}\ \eqref{eq:RRh}\ \text{hold for some}\ (k_t,h_t) \in \mathbb{R}_+^2\right\},
\end{align}
and the social planner's problem has the same form as \eqref{eq:SPref}, replacing $\Gamma$ by $\hat{\Gamma}$ and the return function $F$ with $\hat{F}:\mathbb{R}_+^2 \times \mathbb{R}_+^2 \to \mathbb{R} \cup \{-\infty\}$, defined as
\begin{align}
\label{eq:Fhatdef}
\hat{F}(k_t,h_t,k_{t+1},h_{t+1}):=U\left[Ak_t^\alpha\psi(h_t,h_{t+1})^{1-\alpha}h_t^{1-\alpha+\gamma}+(1-\delta_k)k_t-(1+n)k_{t+1}\right],
\end{align}
for all $(k_t,h_t,k_{t+1},h_{t+1}) \in \mathbb{R}_{++}^2\!\times \mathbb{R}_+^2$ for which $U(\cdot)> -\infty$, and $\hat{F} = F$ if any of $k_t$ or $h_t$ is zero. All other variables for this version of the model are denoted by placing a ``hat'' above the variables introduced in Sections \ref{sec:model} and \ref{sec:equilibrium}, and defined by substituting appropriately $\Gamma$ with $\hat{\Gamma}$ and $F$ with $\hat{F}$.   

As discussed earlier, it is sufficient to show that $\hat{\Gamma}$, $\hat{F}$, and $\beta$ satisfy assumptions \ref{itm:A1} to \ref{itm:A6} to prove the existence of an optimal equilibrium with externalities and the continuity of the value function. However, the presence of increasing returns may introduce nonconvexities in the feasibility correspondence, so the size of the externality, represented by $\gamma$, must be limited for \ref{itm:A6} to hold. This problem can be solved using a different assumption, as shown below.  

The arguments for \ref{itm:A1} and \ref{itm:A5} are nearly identical to those given in the proofs of Lemma \ref{lem:Pidef} and Proposition \ref{prp:bellman}, respectively. To verify \ref{itm:A2}, suppose that $(k',h') \in \hat{\Gamma}(k,h)$ for some $k,h \geq 0$. Let $\omega:=\alpha/(1+\gamma)$. Applying the weighted GM-AM inequality \eqref{eq:WAMGMI} and Jensen's inequality for convex functions to the right-hand side of \eqref{eq:RRkex} yields
\begin{align*}
k' &\leq \frac{A k^\alpha h^{1-\alpha+\gamma}+(1-\delta_k)k}{(1+n)},\\[5pt]
   &\leq \frac{A\left[\omega k + (1-\omega)h\right]^{1+\gamma} +(1-\delta_k)k}{(1+n)},\\[5pt]   
&\leq \frac{A\left[\omega k + (1-\omega)h\right] + (1-\delta_k)k}{(1+n)},\\[5pt]
&= \frac{\left[\omega A + (1-\delta_k)\right]k + (1-\omega)A h}{(1+n)}.
\end{align*}
This implies that $k' \leq \hat{\xi}\|k,h\|$, with 
\begin{equation*}
\hat{\xi}:=\max\left\{\frac{\omega A +(1-\delta_k)}{(1+n)},\frac{(1-\omega)A}{(1+n)}\right\}.
\end{equation*}
Note the similarity of $\hat{\xi}$ with the expression for $\xi$ in \eqref{eq:xidef}. Since $0 < \omega < 1$, the remaining arguments given in the proof of Lemma \ref{lem:Pidef} are still valid, replacing $\alpha$ with $\omega$. The same is true for \ref{itm:A3} and \ref{itm:A4} with this substitution. Given that $\omega \leq \alpha$, it is clear from \eqref{eq:betacond} that the upper bound of $\beta$ for \ref{itm:A4} to hold increases in the presence of externalities, so it becomes a less stringent condition.     

For assumption \ref{itm:A6}, let $(k',h') \in \hat{\Gamma}(k,h)$ and $\lambda \in (0,1]$. Multiplying all inequalities in \eqref{eq:RRkex} by $\lambda$, it follows that $(1-\delta_k)/(1+n)(\lambda k) \leq \lambda k'$ and
\begin{align*}
\lambda k' &\leq \frac{\lambda^{-\gamma}A(\lambda k)^\alpha(\lambda h)^{1-\alpha+\gamma}+(1-\delta_k)(\lambda k)}{(1+n)} \leq \frac{A(\lambda k)^\alpha(\lambda h)^{1-\alpha+\gamma}+(1-\delta_k)(\lambda k)}{(1+n)}, 
\end{align*}   
provided that $\gamma \leq 1$, while the relevant condition for $h'$ follows directly from \eqref{eq:RRh}. Hence $(\lambda k',\lambda h') \in \hat{\Gamma}(\lambda k,\lambda h)$ and part (a) of \ref{itm:A6} holds. To verify \ref{itm:A6}(b), let 
\begin{align}
\label{eq:fhatdef}
\hat{f}(k,h,k',h'):= Ak^\alpha\psi(h,h')^{1-\alpha}h^{1-\alpha+\gamma}+(1-\delta_k)k-(1+n)k'.
\end{align}
Therefore, given that $\lambda \in (0,1]$,
\begin{align*}
\hat{f}(\lambda k,\lambda h,\lambda k',\lambda h')&=A(\lambda k)^\alpha\psi(\lambda h,\lambda h')^{1-\alpha}(\lambda h)^{1-\alpha+\gamma} +(1-\delta_k)(\lambda k)-(1+n)(\lambda k'),\\[5pt]
&=\lambda^{1+\gamma}A k^\alpha\psi(h,h')^{1-\alpha}h^{1-\alpha+\gamma} + \lambda[(1-\delta_k)k-(1+n)k'],\\[5pt]
&\geq \lambda^{1+\gamma}[Ak^\alpha\psi(h,h')^{1-\alpha}h^{1-\alpha+\gamma}+
(1-\delta_k)k-(1+n)k'],\\[5pt]
&=\lambda^{1+\gamma}\hat{f}(k,h,k',h'),
\end{align*}
It follows that $\hat{F}= U \circ \hat{f}$, and
\begin{align*}
\hat{F}(\lambda k,\lambda h,\lambda k',\lambda h') &= \frac{\hat{f}(\lambda k,\lambda h,\lambda k',\lambda h')^\theta-1}{\theta},\\[5pt]
&\geq \frac{\lambda^{(1+\gamma)\theta}\hat{f}(k,h,k',h')^\theta-1}{\theta},\\[5pt]
&= \frac{\lambda^{(1+\gamma)\theta}[\hat{f}(k,h,k',h')^\theta-1]+\lambda^{(1+\gamma)\theta}-1}{\theta},\\[5pt]
&= \lambda^{(1+\gamma)\theta}\hat{F}(k,h,k',h')+U(\lambda^{1+\gamma}).
\end{align*}
This immediately implies that $\hat{F}(\lambda k,\lambda h,\lambda k',\lambda h') \geq \hat{\Phi}_1(\lambda)\hat{F}(k,h,k',h')+\hat{\Phi}_2(\lambda)$, where the functions $\hat{\Phi}_1:(0,1]\to \mathbb{R}$ and $\hat{\Phi}_2:(0,1]\to \mathbb{R}$ are given by
\begin{align*}
\hat{\Phi}_1(\lambda):=\lambda^{(1+\gamma)\theta} \quad \text{and} \quad 
\hat{\Phi}_2(\lambda):=U(\lambda^{1+\gamma}),
\end{align*}
respectively, and are both continuous with $\hat{\Phi}_1(1)=1$ and $\hat{\Phi}_2(1)=0$, as desired. Finally, part (c) of this assumption can be easily verified along the lines of the proof of Proposition \ref{prp:bellman}. The above analysis is summarized in the next proposition.

\begin{prop}
\label{prp:extension}
Assume that $\hat{\Gamma}$, $\hat{F}$, and $\beta$ satisfy \ref{itm:A1} to  \ref{itm:A6} with $\gamma \leq 1$. Then, there exists an optimal equilibrium with externalities for any $(k_0,h_0) \in \mathbb{R}_+^2$. The value function $\hat{V}:\mathbb{R}_+^2 \to \mathbb{R} \cup \{-\infty\}$ satisfies the Bellman equation and is the unique solution to this equation in the space $S$. Moreover, $\hat{V}$ is continuous for any $(\mathbf{k},\mathbf{h}) \in \hat{\Pi}'(k_0,h_0)$, and if $\hat{V}(k_0,h_0) = - \infty$, then for any sequence $\{(k_n,h_n)\}_{n=0}^{+\infty}$ that converges to $(k_0,h_0)$, if follows that $\displaystyle\lim_{n \to +\infty} \hat{V}(k_n,h_n)=-\infty$.
\end{prop}

It is worth pointing out that the previous result is not as restrictive as it seems: without assumption \ref{itm:A6}, an optimal equilibrium with externalities exists, and the value function is upper semicontinuous, independently of the size of $\gamma$. Continuity of the value function, in the sense of the results given in Propositions \ref{prp:bellman}(c) and \ref{prp:extension}, can be obtained for all $\gamma \geq 0$ from \ref{itm:A1} to \ref{itm:A5}, the result of Lemma \ref{lem:pitilde}, and the following assumption:
\begin{enumerate}[label=(A7)]
\item\label{itm:A7} $\hat{\Gamma}(0,0)=\{0,0\}$, $\hat{F}(0,0,0,0)=-\infty$, and there exists a continuous function $\hat{\Psi}:\mathbb{R}_+^2 \to \mathbb{R} \cup \{-\infty\}$ that satisfies $\hat{V}(k_0,h_0) \geq \hat{\Psi}(k_0,h_0)$, for all $(k_0,h_0) \in \mathbb{R}_+^2$, and
 $\displaystyle\lim_{t \to +\infty} \beta^t\,\hat{\Psi}(k_t,h_t) = 0$, for all $(\mathbf{k},\mathbf{h}) \in \hat{\Pi}'(k_0,h_0)$.
\end{enumerate} 
However, proving the existence of such function $\hat{\Psi}$ may not be an easy task, even in the absence of externalities or a simpler version of the model (e.g., without physical capital). See, for instance, \cite{levanmorhaim02}, where they verify this assumption, with some additional technical conditions, for the cases of a standard growth model with logarithmic utility (Example 7.5), a one-sector AK model (Example 7.6), or a very simplified growth model with increasing returns (Example 7.7). In the discussion that follows, it will be shown that \ref{itm:A7} holds for our extended model, hence the continuity of the value function, using a simple monotonicity property possessed by the family of isoelastic utilities defined in \eqref{eq:Udef}.    

\section{Discussion}
\label{sec:discuss}

We emphasized in Section \ref{sec:equilibrium} the importance of using certain functional forms for preferences and technology and their relation to some inequalities for means in  simplifying the proofs. Another relevant aspect, which is often overlooked in the literature, is related to the family of isoelastic utility functions \eqref{eq:Udef}. As indicated by their name, these functions are characterized by having a constant elasticity of substitution and can be obtained as the solution of a boundary value problem, which is shown in Appendix \ref{sec:apputility}. 

\begin{figure}[h!]
\centering
\includegraphics[scale=0.55]{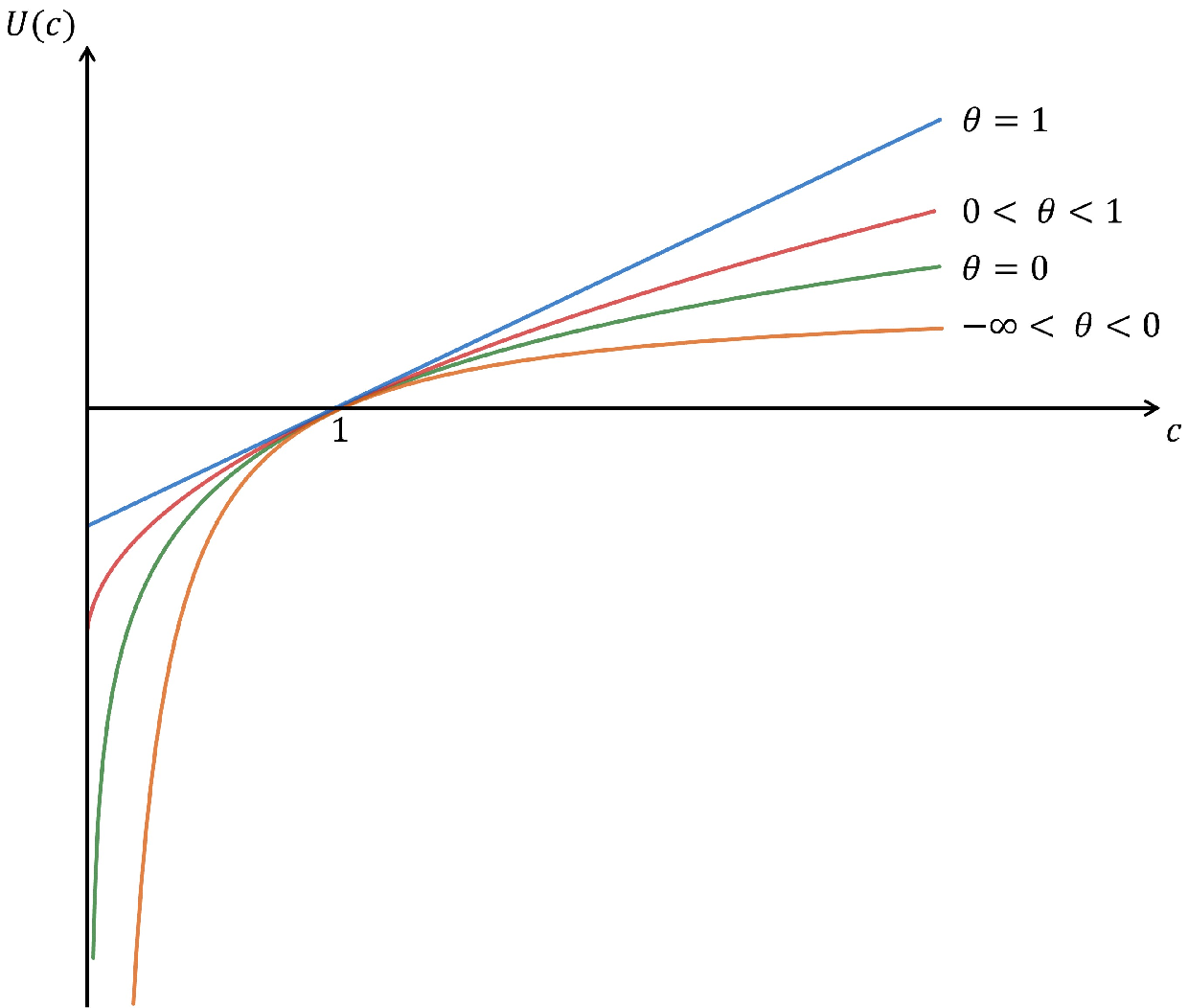}
\caption{The family of isoelastic utility functions $U(c)=(c^\theta-1)/\theta$}
\label{fig:utility}
\end{figure}

A few examples of functions within this class are depicted in Figure \ref{fig:utility}. It is apparent that their curvature depends on the parameter $\theta$, but also that -- taken as a \emph{family} of functions -- all of them are (pointwise) bounded above by $U(c)=c-1$, as $\theta \to 1^-$. Furthermore, they satisfy a parametric continuity with respect to $\theta$, as well as a monotonicity property shown in Proposition \ref{prp:thetamonot} of Appendix  \ref{sec:apputility}. 

For different reasons, it is common use in economics to disregard the additive  constant in \eqref{eq:Udef}, and adopt the alternative formulation
\begin{align}
\label{eq:altutility}
\tilde{U}(c)= 
\begin{cases}
\dfrac{c^\theta}{\theta}, &\text{if}\ -\infty < \theta \leq 1,\ \theta \neq 0,\\[7pt]
\log{c}, &\text{if}\ \theta = 0,
\end{cases}
\tag{\ref{eq:Udef}$'$}
\end{align}      
that preserves the isoelastic property and makes the function homogeneous of degree $\theta$, for all $\theta \neq 0$. This alternative formulation \eqref{eq:altutility} has also the following properties which may prove useful: (i) if $0 < \theta < 1$, then $\tilde{U}(c)$ takes positive values on $(0,+\infty)$, hence is bounded below; (ii) if $-\infty < \theta < 0$, $\tilde{U}(c)$ takes negative values on $(0,+\infty)$ and is unbounded below, but bounded above by zero; and (iii) if $\theta = 0$, as the log function is unbounded and takes both positive and negative values, this case is treated as an ``intermediate'' case between the previous two, with the addition of being analytically tractable. 

Adopting the alternative ad hoc formulation $\tilde{U}$ allows to define spaces for the corresponding return function with convenient properties, that also translate to the value function, but requires different tools for the three cases outlined above, as in \citet{alvarezstokey98}. The unifying approach developed by  \citet{levanmorhaim02} offers a solution to this problem, but at the expense of including topological assumptions that may be difficult to verify in particular cases. This paper shows that using a \emph{family} of isoelastic utility functions is not only closer in spirit to a unified methodology, but also have technical advantages that simplify some of the proofs. To illustrate this point, this section finishes by showing that a parametric monotonicity property satisfied by this family of functions is helpful in using assumption \ref{itm:A7} to prove the existence of an optimal equilibrium with externalities and the continuity of the value function. 

\begin{prop}
\label{prp:discussion}
Assume that $\hat{\Gamma}$, $\hat{F}$, and $\beta$ satisfy \ref{itm:A1} to  \ref{itm:A5}, and \ref{itm:A7}. Then, the results of Proposition \ref{prp:extension} are also true.
\end{prop}

\begin{proof}
Since the return function $\hat{F}$ is bounded below by $-\theta^{-1}$, for all $0 < \theta \leq 1$, it suffices to show that \ref{itm:A7} holds for $-\infty < \theta \leq 0$. Then, the results follow from Lemma \ref{lem:pitilde}, and Proposition 1 and Theorem 3(iii) in \cite{levanmorhaim02}. 

Both $\hat{\Gamma}(0,0)=\{0,0\}$ and $\hat{F}(0,0,0,0) = -\infty$ follow straightforward from the definitions, given in \eqref{eq:Gammaexdef} and \eqref{eq:Fhatdef}, respectively.  Let $(k_0,h_0) \in \mathbb{R}_{++}^2$. It has been established in Lemma \ref{lem:pitilde} that $\Pi'(k_0,h_0)$ is nonempty for all $(k_0,h_0)$ in this set. Since the expressions for $\underline{v}$ and $\overline{u}$ used in the proof of the Lemma remain unchanged for the model with externalities, as can be seen from \eqref{eq:hhatprime} and \eqref{eq:psirhodef}, it is easily verified that the result extends to this case. Hence, the choice of $(\mathbf{k},\mathbf{h}) \in \hat{\Pi}'(k_0,h_0)$ is possible. This immediately implies that 
\begin{align}
\label{eq:Jhatlb}
\hat{J}(\mathbf{k},\mathbf{h}):= \sum_{t=0}^{+\infty} \beta^t \hat{F}(k_t,h_t,k_{t+1},h_{t+1}) > -\infty.
\end{align}
Therefore, $c_t > 0$, for all $t$. On the other hand, from Proposition \ref{prp:thetamonot} in Appendix \ref{sec:apputility}, it is possible to choose $\varepsilon > 0$, so that
\begin{align*}
\hat{F}(k_t,h_t,k_{t+1},h_{t+1})=\frac{\hat{f}(k_t,h_t,k_{t+1},h_{t+1})^\theta-1}{\theta} \geq \frac{\hat{f}(k_t,h_t,k_{t+1},h_{t+1})^{\theta-\varepsilon}-1}{(\theta-\varepsilon)},
\end{align*}
for all $(k_t,h_t)$ in $(\mathbf{k},\mathbf{h})$, where $\hat{f}$ is defined in \eqref{eq:fhatdef}. Let $\hat{\Psi}$ be given by
\begin{align*}
\hat{\Psi}(k_0,h_0):=\sum_{t=0}^{+\infty} \beta^t\frac{\hat{f}(k_t,h_t,k_{t+1},h_{t+1})^{\theta-\varepsilon}-1}{(\theta-\varepsilon)}. 
\end{align*}
Clearly, $\hat{J}(\mathbf{k},\mathbf{h}) \geq \hat{\Psi}(k_0,h_0)$, for all $(\mathbf{k},\mathbf{h}) \in \hat{\Pi}'(k_0,h_0)$, from which it follows that $\hat{V}(k_0,h_0) \geq \hat{\Psi}(k_0,h_0)$, for all $(k_0,h_0) \gg (0,0)$. Given that $\hat{\Psi}$ must be bounded above from assumptions \ref{itm:A2} to  \ref{itm:A4}, and bounded below from \eqref{eq:Jhatlb}, 
\begin{align*}
\lim_{t \to \infty} \beta^t\,\hat{\Psi}(k_t,h_t) =0, \quad \text{for all}\ (\mathbf{k},\mathbf{h}) \in \hat{\Pi}'(k_0,h_0). 
\end{align*}
This concludes the proof.
\end{proof}

\section{Concluding remarks}
\label{sec:conclusion}

This paper studies a discrete-time version of the Lucas-Uzawa optimal growth model  growth in a fairly general setting. Existence of optimal equilibria is proved applying tools of dynamic programming with bounded or unbounded returns, based on the framework by \cite{levanmorhaim02} and \cite{levan06}. The proofs also rely on properties of isoelastic utility and homogeneous production functions and apply well-known inequalities in real analysis, seldom used in the literature, which significantly simplify the task of verifying certain assumptions that are rather technical in nature. Some advantages of adopting a parametric family of isoelastic utility functions, instead of the ad hoc formulation typically used, are also discussed. Directions for future research stemming from our work include determining conditions under which the model exhibits sustained growth, and analyzing the stability of these paths.

\appendix

\section{Some inequalities for means}
\label{sec:inequalities}

\begin{defn}
\label{def:WGM}
Let $x:=(x_1,\ldots,x_n)$ an $n$-tuple of positive real numbers and let $p$ be a real number. If $w:=(w_1,\ldots,w_n)$ is an $n$-tuple of positive weights with $w:=\sum_{i=1}^n w_i$, then the (weighted) \textbf{generalized} or \textbf{power mean} with exponent $p$ is defined as  
\begin{align*}
&M_p(x;w) :=\left(\frac{1}{w}\sum_{i=1}^n w_i x_i^p\right)^{1/p},\text{for all}\ p \neq 0,\ \text{and} \\[5pt] 
&M_0(x;w) :=\left(\prod_{i=1}^n x_i^{w_i}\right)^{1/w}.
\end{align*}
$M_0$ is also known as the weighted \textbf{geometric mean} (GM) and $M_1$ as the weighted \textbf{arithmetic mean} (AM). 
\end{defn}

A general result based on Definition \ref{def:WGM} is called the weighted generalized or power mean inequality.

\begin{thrm} 
\label{thm:WGMI}
Let $x:=(x_1,\ldots,x_n)$ be an $n$-tuple of positive real numbers and let $w:=(w_1,\ldots,w_n)$ be an $n$-tuple of positive weights with $w:=\sum_{i=1}^n w_i$. If $p$ and $q$ are two real numbers such that $p < q$, then  
\begin{align}
\label{eq:WGMI}
M_p(x;w) \leq M_q(x;w),
\end{align}
with equality if and only if $x_1 = x_2 = \cdots = x_n$. 
\end{thrm}

A case of particular interest, which is a direct consequence of Theorem \ref{thm:WGMI}, is the weighted geometric-arithmetic inequality or weighted GM-AM inequality.
\begin{corl} 
\label{cor:WAMGMI}
Let $x:=(x_1,\ldots,x_n)$ be an $n$-tuple of positive real numbers and let $w:=(w_1,\ldots,w_n)$ be an $n$-tuple of positive weights with $w:=\sum_{i=1}^n w_i$. Then  
\begin{align}
\label{eq:WAMGMI}
M_0(x;w)=\left(\prod_{i=1}^n x_i^{w_i}\right)^{\frac{1}{w}} \leq \frac{1}{w}\sum_{i=1}^n w_i x_i=M_1(x;w),
\end{align}
with equality if and only if $x_1 = x_2 = \cdots = x_n$. 
\end{corl}

\begin{rmrk}
\label{rmk:wlog}
Without loss of generality, all the results given in Theorem \ref{thm:WGMI} and Corollary \ref{cor:WAMGMI} can be stated for $w_i \in [0,1]$, $i=1,\ldots,n$, and $w=1$.
\end{rmrk}

Proofs of these results and an in-depth treatment of generalized or power mean inequalities may be found in Chapters 2 and 3 of \cite{bullen03}.

\section{Isoelastic utility functions}
\label{sec:apputility}

\begin{prop}
Let $U \in C^2(0,+\infty)$, with $U' > 0$ and $U'' \leq 0$. The family of isoelastic utility functions \eqref{eq:Udef} is the solution to the following boundary value problem
\begin{align}
\label{eq:BVP}
\begin{cases}
U''(x) + \Theta(x)U'(x) = 0,\\[5pt]
U(1)=0,\ U'(1)=1,
\end{cases}
\end{align}
where $\Theta:(0,+\infty) \to (0,+\infty)$ is given by $\Theta(x)=(1-\theta)/x$ and $\theta$ is a constant such that $-\infty < \theta \leq 1$.
\end{prop}

\begin{proof}
Let $x \in (0,+\infty)$, and rewrite the first equation in \eqref{eq:BVP} as 
\begin{align*}
\frac{U''(x)}{U'(x)} = -\frac{(1-\theta)}{x},
\end{align*}
which is equivalent to
\begin{align*}
\frac{d\log{U'(x)}}{dx} = -\frac{(1-\theta)}{x}.
\end{align*}
Integration over $x$ for $c \in (0,+\infty)$ yields
\begin{align*}
\log{U'(c)}-\log{U'(1)} &= -(1-\theta)\int_1^{\,c} \frac{1}{x}\,dx
= -(1-\theta)\log{c},
\end{align*}
hence, 
\begin{align*}
U'(c)=U'(1)\,c^{-(1-\theta)}. 
\end{align*}
Integrating again both sides of this expression implies that
\begin{align*}
U(c) - U(1) = U'(1)\int_1^{\,c} x^{-(1-\theta)}\,dx = U'(1)\,\frac{c^{\theta}-1}{\theta},
\end{align*}
for $\theta \neq 0$, and
\begin{align*}
U(c) - U(1) = U'(1)\int_1^{\,c} x^{-1}\,dx = U'(1)\log{c},
\end{align*}
if $\theta = 0$.
Therefore,
\begin{align*}
U(c) = 
\begin{cases}
U(1) + U'(1)\,\dfrac{c^\theta-1}{\theta}, & -\infty < \theta \leq 1,\ \theta \neq 0,\\[10pt]
U(1) + U'(1)\log{c}, & \theta = 0,
\end{cases}
\end{align*}
which reduces to \eqref{eq:Udef} after applying the boundary conditions in \eqref{eq:BVP}.
\end{proof}

\begin{prop}
\label{prp:thetamonot}
For each function $U$ in the family of isoelastic utilities \eqref{eq:Udef}, the value of $U(c)$ is decreasing in $\theta$, for all $c \in (0,+\infty)$. 
\end{prop}

\begin{proof}
First, consider the case $-\infty < \theta \leq 0$. We want to show that, for each $c \in (0,+\infty)$, the function $U_c:(-\infty,0] \to \mathbb{R}$, defined by
\begin{align*}
U_c(\theta):=
\begin{cases}
\dfrac{c^\theta-1}{\theta}, &\text{if}\ -\infty < \theta < 0,\\[5pt]
\log{c},                   &\text{if}\ \theta = 0,  
\end{cases}
\end{align*}
decreases with $\theta$. The first derivative of this function is $U_c'(\theta) = 0$, if $\theta = 0$, and 
\begin{align*}
U_c'(\theta)= \frac{c^\theta(\theta \log{c}-1)+1}{\theta^2}, \quad \text{if}\ -\infty < \theta < 0.
\end{align*}
Note that $\sgn{U_c'(\theta)} = \sgn{N_c(\theta)}$, where $N_c(\theta) := c^\theta(\theta \log{c}-1)+1$. Given that $N_c(0)=0$, and 
$N_c'(\theta) = \theta c^\theta(\log{c})^2 \leq 0$, with strict inequality if $c \neq 1$, it follows that $U_c'(\theta) \leq 0$, for all $\theta \in (-\infty,0]$. That is, $U_c$ is decreasing in $\theta$ over this interval. Since $c$ was chosen arbitrarily, this property holds for all $c \in (0,+\infty)$. The proof for $0 < \theta \leq 1$ is completely analogous. 
\end{proof}

\bibliographystyle{elsarticle-harv}
\bibliography{DTEG}

\end{document}